\documentclass[12pt,preprint]{aastex}

\begin{document}

\title{Shrinking Galaxy Disks with Fountain-Driven Accretion from the Halo}

\author{Bruce G. Elmegreen}
\affil{IBM Research Division, T.J. Watson Research Center, 1101 Kitchawan Road,
Yorktown Heights, NY 10598, bge@watson.ibm.com}

\author{Curtis Struck}

\affil{Department of Physics and Astronomy, Iowa State University, Ames, IA
50011, curt@iastate.edu}

\author{Deidre A. Hunter}

\affil{Lowell Observatory, 1400 West Mars Hill Road, Flagstaff, Arizona
86001, dah@lowell.edu}

\begin{abstract}
Star formation in most galaxies requires cosmic gas accretion because the gas
consumption time is short compared to the Hubble time. This accretion
presumably comes from a combination of infalling satellite debris, cold
flows, and condensation of hot halo gas at the cool disk interface, perhaps
aided by a galactic fountain. In general, the accretion will have a different
specific angular momentum than the part of the disk that receives it, even if
the gas comes from the nearby halo. Then the gas disk expands or shrinks over
time. Here we show that condensation of halo gas at a rate proportional to
the star formation rate in the fountain model will preserve an initial shape,
such as an exponential, with a shrinking scale length, leaving behind a
stellar disk with a slightly steeper profile of younger stars near the
center. This process is slow for most galaxies, producing imperceptible
radial speeds, and it may be dominated by other torques, but it could be
important for Blue Compact Dwarfs, which tend to have large, irregular gas
reservoirs and steep blue profiles in their inner stellar disks.
\end{abstract}


\section{Introduction}

Disk accretion is attributed to cold cosmic flows \citep{keres05, dekel09,
van12} and dwarf satellite capture \citep{fox14} through the virial radius,
which may be 20 times larger than the disk \citep[see reviews
in][]{sancisi08,sanchez14}. It is not clear whether these flows reach the
disk directly. \cite{bland07} and \cite{fox14} suggest that the Magellanic
Stream breaks apart and gets ionized by hydrodynamic instabilities without
falling onto the disk. \cite{joung12} also found substantial heating and
ionization of the layers surrounding a cosmic cold flow in simulations of the
Milky Way. The numerical algorithm used for the simulation might also be
important; the AREPO code \citep{nelson13} has such high resolution that the
infalling gas breaks apart by instabilities and heats up before reaching the
disk. These predictions are consistent with the observation that the inflow
rate of neutral clouds around the Milky Way is small, $0.08\;M_\odot$
yr$^{-1}$ \citep{putman12a}, which is too low to maintain the star formation
rate of several $M_\odot$ yr$^{-1}$ \citep{chomiuk11} if that is the only gas
that comes in. Low accretion leads to a relatively rapid quenching of star
formation, contradicting the observation that most disk galaxies form stars
over a high fraction of the Hubble time \citep[e.g.,][]{mannucci10}.

Cosmic simulations suggest that the accreting gas moves toward larger radii over time as
the specific angular momentum of the remaining cosmic filament increases; this leads to the
popular picture of inside-out star formation \citep{pichon11}. Observations agree that
accretion occurs today in the outer parts of galaxies \citep{lemonias11,moffett12,moran12}.
This causes a problem if most of the star formation takes place in the main disk, much
further inward. Why does the star formation rate equal the accretion rate, as analytical
models suggest \citep{bouche10,dave12,lilly13,dekel13}, if these rates occur in different
locations? One needs in-plane accretion at a comparable rate to deliver the fresh outer gas
to the inner star-forming regions. \cite{dekel13} discuss how this might happen using a
model based on energy dissipation. \cite{forbes14} consider in-plane accretion that is
driven by disk torques from interacting clumps and turbulent viscosity.

Another model of disk accretion assumes that cosmic gas goes first into a hot
halo and then condenses onto the disk in a second step via thermal cooling at
the interface. A galactic fountain could provide the cool surface needed for
this condensation \citep{fraternali06, marinacci10,
marinacci11,putman12b,hobbs13}. The fountain model avoids the question of
whether the cosmic gas goes directly from the virial radius to the disk, and
it avoids the problem of accretion onto the outer regions when the gas is
needed in the inner regions to form stars. Cosmic gas may be deposited in the
far outer regions by cold streams while the inner disk is fed by condensing
halo gas immediately adjacent to it. \cite{fraternali14} summarize the
advantages of this model, including many comparisons to observations
\citep{marasco12,marasco13}. Halo condensation gives an accretion rate that
is approximately proportional to the star formation rate at the same
position, and therefore to the existing column density of molecular or
self-gravitating gas. Such a model sustains star formation in the main disk.

Cosmic accretion should generally have a different specific angular momentum
than the part of the disk that receives it. There will be some generation of
turbulence to dissipate the relative kinetic energy, and an inflow or outflow
of the combined gas until radial force equilibrium is achieved in a new
circular orbit. Even in the fountain model, where the accreting gas in the
disk comes directly from the adjacent rotating halo, the high velocity
dispersion in the halo and the resulting radial pressure gradient cause it to
rotate more slowly than the cold disk beneath it. The result will be a
gradual shrinkage of the gas disk as it forms stars from condensing halo gas.

This paper calculates the radial drift speed from halo condensation at the
local star formation rate in order to determine how such a drift affects the
radial profiles of the gaseous and stellar components of the disk. The
results suggest a relatively long timescale for disk evolution forced this
way, so the process may not have strong observational signatures in disks
dominated by other torques, such as bars and spirals. However it may be
evident in dwarf galaxies without these stellar structures. Dwarf galaxy star
formation also seems to proceed from the outside-in \citep{zhang12}, as if
following some inflowing gas, unlike the case in spiral galaxies. The star
formation history in the Large Magellanic Cloud shows an outside-in trend
too, although interactions with the Small Magellanic Cloud could be involved
\citep{dobbie14}.

Other processes involving extra-planar gas should stimulate the same type of
halo condensation and disk shrinkage. \cite{struck09} suggested that gas at
the disk-halo interface follows the motions of spirals and bars, which
concentrate star formation and increase the turbulent pressure locally. They
model the resulting fountain and explain the vertical rotational velocity
gradient that is observed in the interface region. \cite{kim06} simulated the
generation of turbulence by instabilities at high latitude above spiral
density wave shocks; turbulent structures like this could enhance halo
condensation too.

In what follows, the equations for disk evolution driven by accretion of low
angular momentum gas are solved for the case where the accretion rate is
proportional to the local star formation rate. Solutions are shown that
include depletion of gas by star formation and disk winds, and with rates for
accretion and star formation that change over cosmological time. We show that
the star-forming part of the gas disk shrinks in a self-similar way
throughout this process, which means that an initially exponential profile
for this component stays exponential with a shrinking scale length. This is
important because exponential profiles appear very early in the universe
\citep{fathi12}, perhaps as a result of clumpy disk dynamics
\citep{bournaud07,elmegreen13}, and subsequent accretion at the star
formation rate should not distort this profile too much.  Specific
applications to Blue Compact Dwarfs are discussed in Section
\ref{bcdsection}.

\section{Model: Disk evolution with local halo accretion}
\subsection{Surface Density Evolution}

We assume a that galactic disk accretes locally at a rate equal to $\dot
\Sigma_{\rm acc}$ in units of mass per area per time. In the fountain model,
this accretion is affected primarily by the star-forming component of the
disk, so we divide the gas column density into two components, $\Sigma$ for
the star forming part, such as the molecular and self-gravitating clouds, and
$\Sigma_{\rm diff}$ for the non-star-forming part, such as the atomic diffuse
gas. These two components interchange mass and mix with accreted gas on
dynamical timescales, but only the fraction of the total that is involved
with star-formation will drive the accretion and get depleted as stars form.

With this assumption, the star formation rate is $\Sigma_{\rm
SFR}=\Sigma/\tau_{\rm SF}$, with a fraction $R$ of the forming stellar mass
quickly returned to the gas in the form of supernovae and stellar winds. In
general, there is also a disk wind. A common approximation is to use a mass
loss rate in the wind proportional to the star formation rate,
$w\Sigma/\tau_{\rm SF}$, for constant $w$
\citep[e.g.,][]{edmunds90,bouche10,dave12,lilly13}. Then the continuity
equation for gas surface density, including radial drift speeds from angular
momentum redistribution, $v_{\rm r}$ and $v_{\rm r,diff}$ for the
star-forming and non-star-forming parts, is
\begin{equation}
{{\partial \left(\Sigma+\Sigma_{\rm diff}\right)
\over{\partial t}}}+{1\over r}{{\partial}\over{\partial r}} r \left(\Sigma v_{\rm r}+
\Sigma_{\rm diff}v_{\rm r,diff}\right) = {\dot \Sigma_{\rm acc}}
-{{1-R+w}\over{\tau_{\rm SF}}}\Sigma .
\label{cont2}
\end{equation}

We also assume that cosmic accretion has some rotation speed, $v_{\rm acc}$,
that is not generally the same as the circular speed of gas that is already
in the disk at that radius, $v_{\rm circ}$. The equation for conservation of
angular momentum for the two components is then
\begin{equation}
{{\partial \left(\Sigma+\Sigma_{\rm diff}\right)
r v_{\rm circ}}\over{\partial t}} + {1\over r}{{\partial}\over{\partial r}}
r^2 v_{\rm circ} \left(\Sigma v_{\rm r}+
\Sigma_{\rm diff}v_{\rm r,diff}\right) = {\dot\Sigma_{\rm acc}}rv_{\rm acc}
-{{1-R+w}\over{\tau_{\rm SF}}}\Sigma rv_{\rm circ} .
\label{ang}
\end{equation}
The left-hand side of this equation is the time derivative of the local
disk angular momentum, following the radial drift, and the right-hand side
is the angular momentum added to this part of the disk.

Assuming the circular speed changes only slowly with time, we separate
$\partial \left(\Sigma+\Sigma_{\rm diff}\right)/\partial t$ from the first
term on the left of equation (\ref{ang}) and divide the whole equation by
$rv_{\rm circ}$. A term like $(r/v_{\rm circ})(dv_{\rm circ}/dr)$ enters so
we assume a power law rotation curve of the form $v_{\rm circ}\propto
r^\beta$, which makes this term equal to $\beta$. With these modifications to
equation (\ref{ang}) we subtract equation (\ref{cont2}) to eliminate
$\partial\left(\Sigma+\Sigma_{\rm diff}\right)/\partial t$, several common
geometric terms, and the wind and star formation terms. The result is
\begin{equation}
{{v_{\rm r}(1+\beta)\Sigma\over{r}}}=
-{\dot \Sigma_{\rm acc}}\left(1-{{v_{\rm acc}}\over {v_{\rm circ}}}\right).
\label{vrequation}
\end{equation}
for the star-forming component. We also get $v_{\rm r,diff}=0$ for the non-star forming
component, leaving $\Sigma_{\rm diff}$ unchanged, because it has no source terms on the
right-hand sides of equations (\ref{cont2}) and (\ref{ang}).

This result is consistent with the finding by \cite{bouche10}, \cite{dekel13}
and others that the atomic phase of interstellar gas has approximately
constant mass over time with the accretion feeding only the star formation
component. Our modification here is to consider that the accreted gas has a
lower specific angular momentum than the disk gas, in which case the fraction
of the disk mass involved with star formation moves inward to conserve total
angular momentum (eq. \ref{vrequation}). This result is also consistent with
the observation that the radial distribution of HI is usually much flatter
than the molecular distribution, which matches that of the star formation
rate and approximately also that of the underlying disk stars.

We assume that ${\dot \Sigma_{\rm acc}}$ is proportional to the column
density of disk gas that is involved with star formation,
\begin{equation}
{\dot \Sigma_{\rm acc}}=A\Sigma,
\label{sigmaacc}
\end{equation}
because of the galactic fountain process described in the introduction,
along with a Kennicutt-Schmidt relation that is linear between the star
formation rate and the dense component of the gas column density
\citep{kenn12}.

In the model by \cite{fraternali14} and collaborators, the accreted gas is
from a large reservoir in the hot halo, in which case it was formerly in
hydrostatic equilibrium in the potential well of the galaxy and did not
have much rotational support. Observations indicate a $\sim10$\%-20\%
slower rotation of the disk-halo interface than the disk itself
\citep[e.g.,][]{marinacci11}. This disk-halo interface is where the
condensation of halo gas onto the fountain is presumably happening, so the
rotation speed there is already a mixture of the disk speed from supernova
and wind ejecta, and the halo speed from hot gas condensing onto it. The
unseen halo speed is probably even lower than the observed interface speed.
For simplicity, we write,
\begin{equation}
{{v_{\rm acc}}\over{v_{\rm circ}}}\sim{\rm constant}<1.
\label{vacc}
\end{equation}
In what follows, this ratio will only be compared to unity, so slightly
different radial variations in $v_{\rm acc}$ and $v_{\rm circ}$ will not
matter much.

The solution to equation (\ref{vrequation}) is now
\begin{equation}
v_{\rm r}=-{{Ar}\over{(1+\beta)}}\left(1-{{v_{\rm acc}}\over{v_{\rm circ}}}\right)
\equiv -\gamma r.
\label{vrsol}
\end{equation}
We interpret this equation to mean that the star-forming part of the gas
moves in at the speed $v_{\rm r}$.  This does not mean that individual
molecular clouds are drifting at that speed. As mentioned above, there should
be a continuous mixing between the star-forming and diffuse parts following
star formation, stellar gas return, molecular cloud disruption, and so on.
Thus, equation (\ref{vrsol}) may also be interpreted to mean there is an
inward drift of the whole interstellar medium at the speed $v_{\rm
r}\Sigma/(\Sigma+\Sigma_{\rm diff})$, driven by the star-forming part through
its interaction with the slower-moving halo. If accretion enables star
formation at about the same rate, and star formation drives accretion as in
the fountain model, then ${\dot\Sigma}_{\rm acc}=A\Sigma=A\tau_{\rm
sf}\Sigma_{\rm SFR}$, giving $A\approx 1/\tau_{\rm SF}$.  Thus, we can
equivalently interpret our equations to mean that the star formation rate
scales linearly with the fraction $\Sigma/(\Sigma+\Sigma_{\rm diff})$ of the
total gas, and this fraction drifts at $v_{\rm r}$.

The linear dependence of $v_{\rm r}$ on $r$ implies a self similar shrinking.
To solve for the resulting $\Sigma(r,t)$, we substitute $v_{\rm r}$ into
equation (\ref{cont2}) and simplify to get
\begin{equation}
{{\partial \Sigma}\over{\partial t}}=
\left(2\gamma+A-{{1-R+w}\over{\tau_{\rm SF}}}\right)\Sigma+\gamma r
{{\partial}\over{\partial r}} \Sigma.
\label{sigmadot}
\end{equation}
In the discussion above, the entire coefficient of the $\Sigma$ term on the
right is independent of radius but not necessarily time, so we represent it
by
\begin{equation}{\cal A}(t)=2\gamma+A-{{1-R+w}\over{\tau_{\rm SF}}};
\label{cala}\end{equation}
then we have an evolution equation of the form
\begin{equation}{\dot \Sigma}={\cal A}\Sigma+\gamma r\Sigma^{\prime}\end{equation}
for dot and prime representing time and radial derivatives. The first term on
the right can be eliminated by substituting
$\Sigma(r,t)=\Sigma_0(r,t)\exp(\int_0^t {\cal A}dt)$. Then
\begin{equation}{\dot \Sigma_0}=r\gamma\Sigma_0^{\prime}.\end{equation}
This equation is solved by setting $\Sigma_0(r,t)$ equal to any function
$\chi(y)$ of the variable $y=re^{\gamma t}$, because then ${\dot
\Sigma_0}=r\gamma e^{\gamma t} d\chi(y)/dy$ and $\Sigma_0^{\prime}= e^{\gamma
t}d\chi(y)/dy$. Thus the star-forming part of the gas column density becomes

\begin{equation}
\Sigma(r,t)=\chi\left(re^{\gamma t}\right)e^{\int_0^t {\cal
A}dt};\end{equation} $\chi$ represents the radial profile at $t=0$. We
consider an initial profile that is exponential, so $\chi(y)=e^{-y/R_{D0}}$
for $y=r e^{\gamma t}$ and initial scale length $R_{D0}$.  Then
\begin{equation}\Sigma(r,t)=\Sigma(0,0)e^{\int_0^t
{\cal A}dt} e^{-r/R_D(t)}\label{solution}\end{equation} with exponentially
shrinking scale length, $R_D(t)=R_{D0} e^{-\gamma t}$, and initial central
value $\Sigma(0,0)$. Evidently, the accretion makes the central column
density of this disk component increase with time as $e^{\int_0^t {\cal
A}dt}$.

Because we have assumed ${\cal A}$ to be independent of radius, the radial
profile of the star-forming part of the disk is always exponential following
an initial exponential shape, even as it shrinks with the accretion of
low-angular momentum halo gas. We might expect ${\cal A}$ to have some radial
variation, however, even if the star formation rate is locally proportional
to the accretion rate ($A$ is constant), because of the likely radial
variations of the rotation curve slope, $\beta$, and the wind mass loading
factor, $w$ (cf. eqn. \ref{cala}).  The rotation curve should flatten with
increasing $r$, making $\beta$ decrease and $\gamma$ increase. The wind
factor should increase with $r$ as the disk material is less bound to the
galaxy in the outer parts. These two effects compensate each other in ${\cal
A}$. The other parameters, namely, the gas return fraction $R$ and the
differential circular speed, $(1-v_{\rm acc}/v_{\rm circ}),$ should not vary
as much (especially if $v_{\rm acc}/v_{\rm circ}<<1$). Also, $\tau_{\rm sf}$
is not observed to vary with radius in local disk galaxies for the star
forming part of the interstellar medium, i.e., the CO-emitting gas
\citep{schruba11}. Thus ${\cal A}$ may not vary much. The effects of variable
${\cal A}$ can be understood intuitively because the inflow speed $v_r$
scales with this quantity. Locally high ${\cal A}$ can make a trough, for
example, by increasing the inflow rate through that region.

The star-forming gas mass in the disk is
\begin{equation}M(t)=2\pi\int_0^\infty r\Sigma(r,t)dr
= 2\pi R_{\rm D}(t)^2\Sigma(0,t)=2\pi R_{\rm D0}^2
\Sigma(0,0)e^{\int_0^t {\cal A}t},
\end{equation}
showing the same exponential growth.  This type of growth arises because the
accretion rate is assumed to be proportional to the star-forming gas mass
itself, by feedback through star formation, a galactic fountain, and
condensation of halo gas onto the cool fountain debris.

\subsection{Sample Solutions}

The accretion rate will generally change with time because both $\tau_{\rm
sf}$ and the rate of cosmic accretion into the virial radius change. Sample
solutions derived numerically are shown in Figures 1 and 2 for two models. On
the left-hand side of each figure, we assume in the first model that the
accretion rate equals the star formation rate, which gives $A=1/\tau_{\rm
sf}$.  Following \cite{dekel13}, we use $\tau_{\rm sf}=0.17t_{\rm H}$, where
$t_{\rm H}$ is the Hubble time. On the right-hand side of each figure, we
assume in the second model that $A$ also scales with the halo accretion rate,
normalized to the halo accretion rate at the present time:
\begin{equation}
A={1\over{0.17t_{\rm H}}}\times{{{\dot M}_{\rm halo}}\over{{\dot M}_{\rm halo,z=0}}}
\label{a2}
\end{equation}
where \citep{dekel13}
\begin{equation}
{\dot M}_{halo}=30M_{\rm halo,12}e^{-0.79z}(1+z)^{2.5};
\end{equation}
$M_{\rm halo,12}$ is the mass of the halo in units of $10^{12}\;M_\odot$,
and $z$ is the redshift, related to the expansion time $t_{\rm H}$ by
$\Lambda$CDM cosmology from \cite{planck14} where $H_0=67.3$ km s$^{-1}$
Mpc$^{-1}$ and $\Omega_{\rm m}=0.315$.

For the evaluation of $v_{\rm r}$, we assume $(1-v_{\rm acc}/ v_{\rm
circ})=0.1$ for all radii, so there is a 10\% mismatch between the halo and
disk spins. We also assume a flat rotation curve, $\beta=0$, in Figure 1, and
a solid body rotation curve, $\beta=1$, in Figure 2.  From equation
(\ref{vrsol}), we see that the evolution is slower for higher $\beta$ and
lower $(1-v_{\rm acc}/v_{\rm circ})$.

The starting time for the accretion is taken to correspond to a redshift of
$z=2$, which is in the clumpy disk phase when the smooth exponential disk
starts to appear \citep[e.g.,][]{genzel11}. All of the surface density
profiles are normalized in the figures to the central value of the initial
profile. The curves in the bottom panels show the instantaneous radial
profiles at equally spaced times from the start (green curve) to today (red
curve). The profiles remain exponential as they steepen, as implied by
equation \ref{solution}.

The top panels in the figures show the stellar surface densities calculated
at each radius by integrating the local star formation rate over time,
\begin{equation}
\Sigma_{\rm stars}(t_{\rm H})=(1-R)\int_{t_H=3.27}^{t_H} \Sigma dt/\tau_{\rm sf} ;
\end{equation}
$t_{\rm H}=3.27$ Gyr at the start, where $z=2$, and $t_{\rm H}$ increases
with cosmic time until today, when $t_{\rm H}=13.74$ Gyr. Because $\Sigma$ is
normalized to the central value at the start, the stellar surface density is
not scaled to real units. We assume $R=0.5$ here
\citep[e.g.,][]{edmunds05,leitner11,torrey12}, and for the gas evolution
equation (\ref{sigmadot}), $w=0.2$ \citep{zahid12}. The stellar profiles
steepen with time, but they do not remain exponential like the gas because
the stars have the gas profiles from the times when they formed.

Also in the top panels, plotted with black curves and using the right-hand
axes, are the average, mass-weighted ages of the stellar disks. The average
ages are smaller near the center because the gas has moved in and formed an
increasing proportion of stars there over time.  This is outside-in star
formation.

Surface density evolution by fountain-driven accretion is faster for more
massive galaxies because of the $(1+\beta)$ term in $v_{\rm r}$; massive
galaxies tend to have flatter rotation curves in their bright inner regions
\citep{rubin85}. The age dip at the center is also sharper for higher mass
galaxies, for the same reason. Because we have ignored bulge formation and
other accretion processes that happened earlier, the present results are
only meant to show the part of the evolution from fountain-driven
accretion.

\subsection{A Fit to Blue Compact Dwarfs}
\label{bcdsection}

There could be an application to late-type galaxies. Blue Compact Dwarfs
(BCDs) have high ratios of gas mass to star mass, low metallicities, nearly
solid-body rotation curves, and a generally young appearance.  They also have
steeper and bluer exponential profiles in the inner disk than the outer disk
\citep{hunter04,hunter06,jan14}.  Dwarf Irregulars tend to have bluer inner
regions too, suggesting outside-in star formation \citep{zhang12}. For
example, the far ultraviolet scale lengths for low mass galaxies are about
half the 3.6$\mu$m scale lengths, unlike the case for high-mass galaxies
where these scale lengths are about the same \citep{zhang12}. Low mass
galaxies have weak or no spiral arms to drive a radial evolution, and they
have sustained star formation at a steady, though bursty, rate all of their
lives. If the freshly accreted gas that drives star formation has relatively
low angular momentum, then there could have been a radial accretion like that
discussed here.

Figure 3 shows the radial profiles of stellar mass surface density in four
dwarf galaxies with Type III stellar surface brightness profiles: exponential
profiles with radius that have an abrupt break and become shallower in the
outer disk \citep[see][]{herrmann13}. These profiles are common among BCDs or
dwarf galaxies with significant central star formation (current or past).
Haro 29 and NGC 3738 are classical BCDs with young central parts and older,
smooth outer parts \citep{ashley13}. NGC 1569 is a young starburst
\citep{johnson12}, and DDO 70 is a relatively normal dwarf irregular with
evidence for an event in the past that produced an enhancement in the center
of the galaxy \citep{hunter04,hunter06}. Each galaxy has steeper surface
brightness and stellar mass surface density profiles in the inner disk
compared to the outer disk, presumably because of some accretion that led to
the starburst there. The observed mass surface density profiles are shown as
red curves \citep[from][]{zhang12}, and compared to theoretical models of
stellar surface density calculated as above. The models fitted the inner and
outer slopes primarily by adjusting the accretion history.

All models had the same values of $(1 - v_{\rm acc}/v_{\rm circ}) = 0.5$ (a
relatively large angular momentum mismatch between the halo and the disk),
$\beta = 0.3$ (a slowly rising rotation curve), $(1 - R + w) = 1.4$ (a
relatively strong wind for these small galaxies, i.e., $R = 0.5$ as above
but now $w = 0.9$ for the wind), and a redshift of 2 at the start of the
evolution, as above. The two galaxies with the relatively small increase of
slope in the center, DDO 70 and NGC 1569, assume $A = 1/(0.17t_{\rm H})$
without the cosmic accretion multiplier in equation (\ref{a2}), while the
two galaxies with large increases of slope in the center, Haro 29 and NGC
3738, assume $A = 1/(0.17t_{\rm H})$ up to 1 Gyr ago and then a burst of
accretion toward the end with 10 times this rate. In the figure, the top
blue curve is the one corresponding to the present time, while the other
curves show the evolution in equal time steps.

These simple models fit Haro 29 and DDO 70 very well (Haro 29 has a small
centrally flat part for which we have artificially moved the galaxy profile
slightly to the left in the figure). For NGC 1569 and NGC 3738, the models
fit the outer profiles well, but produce inner mass profiles that are
slightly less massive than what are observed with roughly the same average
slope. Thus, aside from some irregularities in the profiles, fits to the
average inner and outer slopes are possible with the accretion model.
Galaxies with steeper inner profiles have a larger recent influx of low
angular momentum gas. If the influx is steady, then the stars build up with
a steadily decreasing scale length and there is a more gradual transition
between the inner and outer parts.

Also shown in Figure 3 are the gas surface density profiles derived from the
deprojected HI mass surface densities multiplied by 1.36 to account for
Helium and heavy elements.  These profiles go out much further in radius than
the stellar parts shown in the panels below each one. They are plotted in
physical units, $M_\odot$ pc$^{-2}$, using the left-hand axes. The molecular
gas surface density, which presumably follows the star formation rate and
inner stellar disk better, is probably much lower than the HI+He surface
density throughout most of the plotted region. The HI+He profile is not
exponential even though the profile of star formation rate is
\citep[H$\alpha$ profiles are in][]{hunter04}.

\section{Summary}

The interstellar matter in galactic gas disks should migrate inward if the
disk accretes from the halo regions at slightly lower specific angular
momentum. Accretion at the disk-halo interface \citep{fraternali06} could
cause such migration and have a noticeable effect on the stellar age
distribution after several Gyr.  We have shown that this process preserves
the exponential shape of the star-forming part of a gas disk initially in
this form, while shrinking the disk scale length to conserve angular
momentum. The radial drift speed is small, on the order of the disk size
divided by the Hubble time, or $\sim1$ km s$^{-1}$. Gas consumption by star
formation and gas expulsion by disk winds do not change the exponential shape
nor the rate of evolution of the scale length; they only change the absolute
value of the surface density. Fits of the model to BCD galaxies can account
for the steep inner stellar profiles of those galaxies while explaining the
origin of the dense gas in the center that drives the starburst.

We are grateful to the referee for pointing out a more general solution to
equation \ref{sigmadot} than we originally derived.

\clearpage
\begin{figure}\epsscale{.9}
\plotone{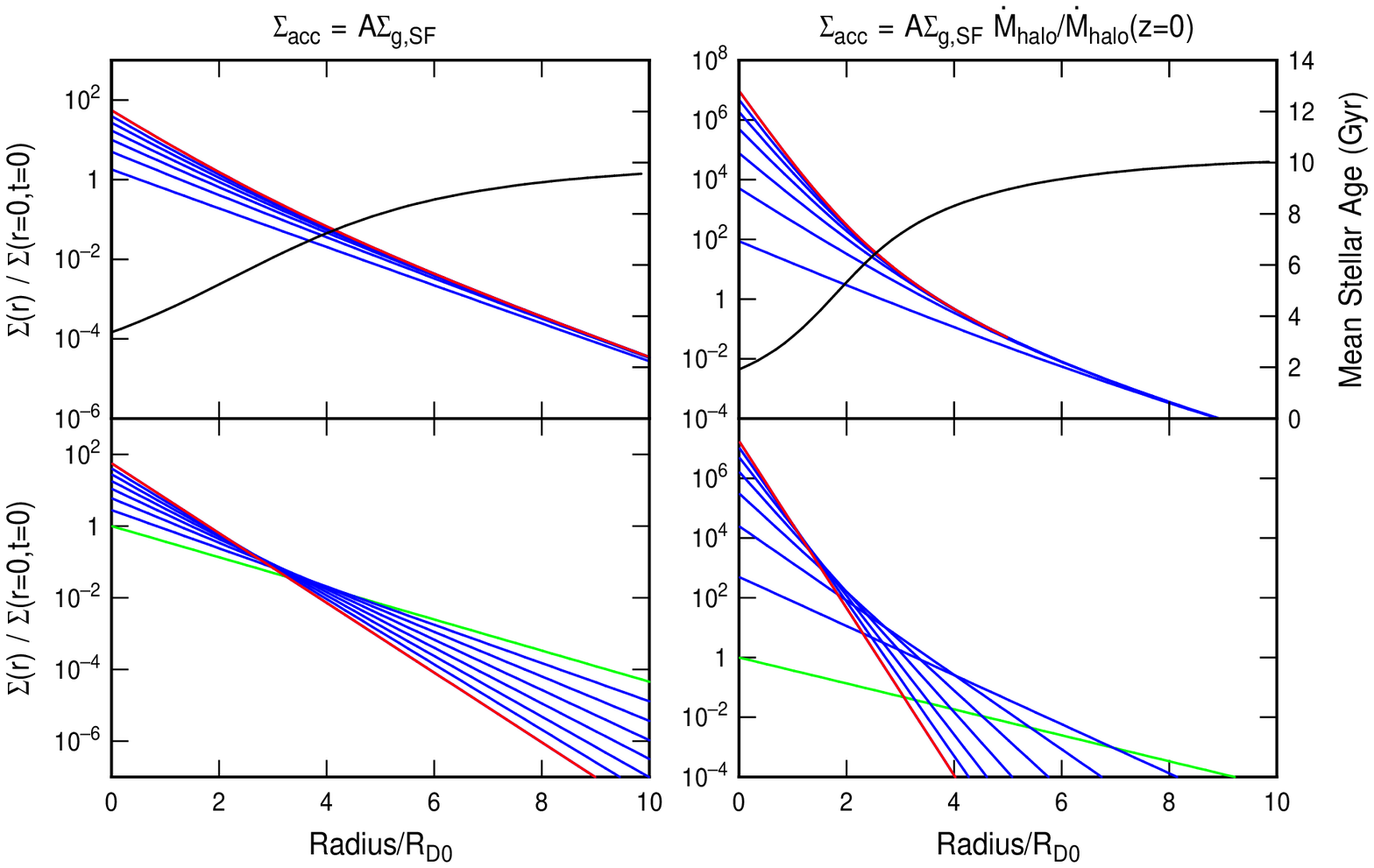}
\caption{Solutions for the gas (bottom) and stellar (top)
surface density profiles as functions of radius (arbitrary vertical scales),
given for equally spaced times between redshift
$z=2$ (green curves) and today (red curves). The left-hand panels are for
an accretion rate onto the disk equal to the local star formation rate, and the
right-hand panels show a more time dependent accretion rate equal to the
local star formation rate multiplied by the halo accretion rate normalized to the
value at $z=0$.
The top panels also show (black curves) the mass-weighted mean stellar ages
using the right-hand axes. This figure is for a model with a flat rotation
curve.} \label{fig1}\end{figure}

\clearpage
\begin{figure}\epsscale{.9}
\plotone{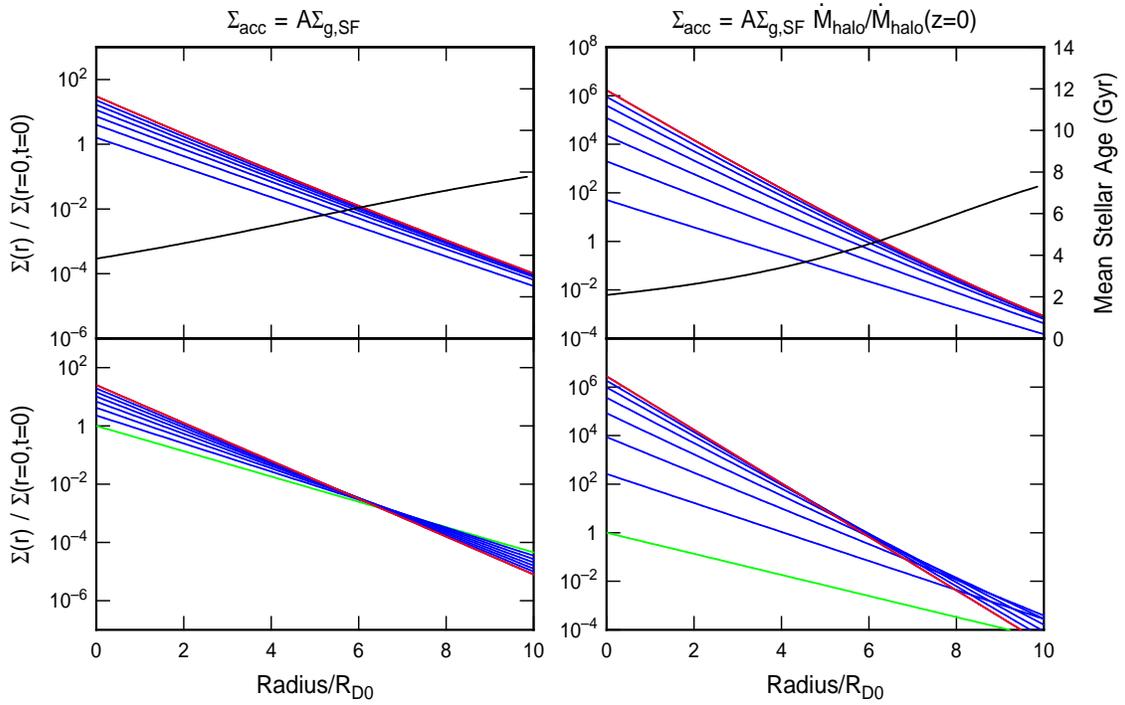}
\caption{Solutions for disk evolution as in Fig. 1 but
with a solid body rotation curve.} \label{fig2}\end{figure}

\clearpage
\begin{figure}\epsscale{.9}
\plotone{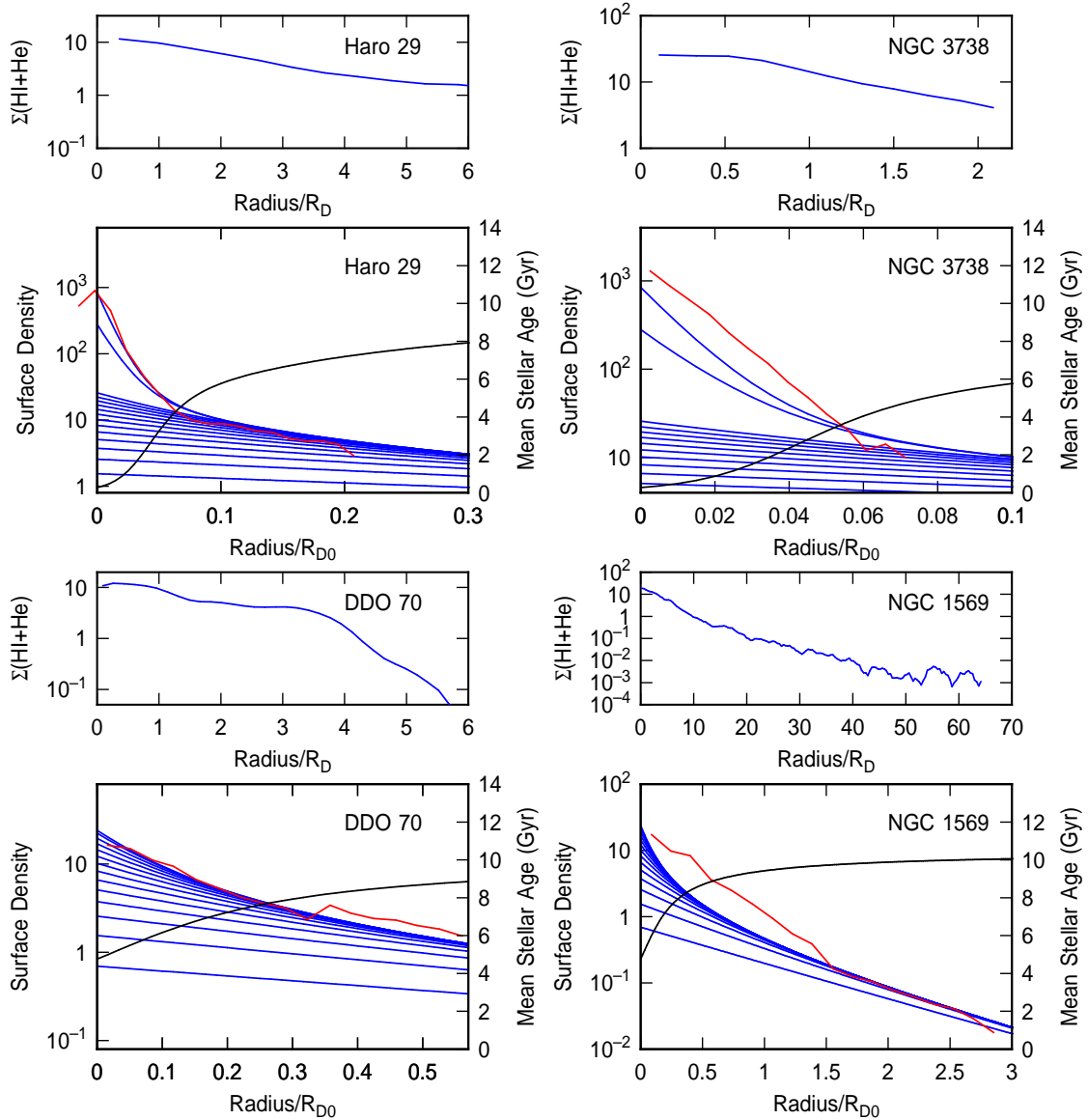}
\caption{A comparison between the mass surface density profiles in four
centrally bursting galaxies (red curves, from Zhang et al. 2012) and models with radial accretion from low
angular momentum cosmic gas (blue curves).  The model curves increase over
time as the stellar disk builds up, and they steepen toward the center
as the gas disk which forms it shrinks.  The radius in the models is normalized to the
scale length of the initial disk, and in the observations it is
varied to match the models. The vertical scale interval is the same for the models and
observations. For the observations, the vertical scale is the real
mass surface density in units of $M_\odot$ pc$^{-2}$, while
for the model, the vertical position has been adjusted to fit the observations.
The black curve in each panel is the mean stellar age, using the scale on the
right-hand axis.  The deprojected HI+He mass surface density profiles are also shown, above
each stellar profile, plotted in physical units of $M_\odot$ pc$^{-2}$ and with an
abscissa equal to the ratio of the radius to the current V-band scale length.  The HI+He
profiles extend much further than the stars in these gas-rich systems and they are
not generally exponential like the star formation rate and stellar mass profiles.}
\label{fig3}\end{figure}

\end{document}